\documentclass[sigconf]{acmart}
 \pdfoutput=1
\AtBeginDocument{%
  \providecommand\BibTeX{{%
    \normalfont B\kern-0.5em{\scshape i\kern-0.25em b}\kern-0.8em\TeX}}}

\acmConference[CHI: Conference on Human Factors in Computing Systems '24]{}{May 11--16,
  2024}{Honolulu, USA}
\setcopyright{none}

\acmConference[1st HEAL Workshop at CHI Conference on Human Factors in Computing Systems]{}{May 12}{Honolulu, HI, USA}

\acmDOI{}

\acmISBN{}

\begin{document}

\title{Exploring Subjectivity for more Human-Centric Assessment of Social Biases in Large Language Models}

\author{Paula Akemi Aoyagui}
\email{paula.aoyagui@mail.utoronto.ca}
\affiliation{%
  \institution{University of Toronto}
  \city{Toronto}
  \country{Canada}
}

\author{Sharon Ferguson}
\email{sharon.ferguson@mail.utoronto.ca}
\affiliation{%
  \institution{University of Toronto}
  \city{Toronto}
  \country{Canada}
}

\author{Anastasia Kuzminykh}
\email{anastasia.kuzminykh@utoronto.ca}
\affiliation{%
  \institution{University of Toronto}
  \city{Toronto}
  \country{Canada}
}


\begin{abstract}
An essential aspect of evaluating Large Language Models (LLMs) is identifying potential biases. This is especially relevant considering the substantial evidence that LLMs can replicate human social biases in their text outputs and further influence stakeholders, potentially amplifying harm to already marginalized individuals and communities. Therefore, recent efforts in bias detection invested in automated benchmarks and objective metrics such as accuracy (i.e., an LLMs output is compared against a predefined ground truth). Nonetheless, social biases can be nuanced, oftentimes subjective and context-dependent, where a situation is open to interpretation and there is no ground truth. While these situations can be difficult for automated evaluation systems to identify, human evaluators could potentially pick up on these nuances. In this paper, we discuss the role of human evaluation and subjective interpretation in augmenting automated processes when identifying biases in LLMs as part of a human-centred approach to evaluate these models.

\end{abstract}

\ccsdesc[500]{Human-centered computing~HCI theory, concepts and models}
\ccsdesc[500]{Human-centered computing~HCI design and evaluation methods}
\ccsdesc[500]{Human-centered computing~Natural language interfaces}

\keywords{Human-AI collaboration, Large Language Models, Bias  Evaluation, Human Evaluation, Social Bias, Gender Bias}


\maketitle

\section{Introduction}
Evaluating biases in Large Language Models (LLMs) is paramount considering not only their widespread use and popularization but also the growing concerns that these models can mirror and amplify harmful biases found in language and society \cite{gehman2020realtoxicityprompts, parrish2021bbq, blodgett2020language, kotek2023gender, sharma2024generative}. In LLMs, social biases manifest, for example, in text outputs that stereotype, misrepresent or use derogatory language against a group or individual based on a characteristic such as race, age, gender, sexual preference, political ideology, religion, etc \cite{dhamala2021bold}. For instance, if a biased LLM powers an automated hiring system designed to screen hob applicants, there is a risk that some candidates might be offered better or worse opportunities (i.e. allocational harm) \cite{ferrara2023should}. Similarly, in automated content moderation tasks, an LLM could miss out on nuances and mistakenly classify text from a specific group as toxic (i.e., representational harm). Furthermore, beyond automated system applications, there are potential risks for human-AI collaboration as well, since past research has pointed humans might \textit{``inherit"} biases from an AI system's recommendations \cite{vicente2023humansinherit}. More specifically, \citet{ferguson2023something} demonstrated how an LLM's text assessment can affect a human's text output in both linguistic characteristics and positionality. Hence, plenty of effort has been spent developing methods to detect and quantify biases in LLMs. 

Bias evaluation in LLMs is most often based on automated methods and benchmarks \cite{gallegos2023bias}, with human evaluation mainly used to verify the results from the automated analysis \cite{esiobu2023robbie, dhamala2021bold}. \citet{chang2023survey}'s survey on LLM evaluation concludes that automated methods are preferred and more popular mainly because \textit{``automatic evaluation does not require intensive human participation, which not only saves time but also reduces the impact of human subjective factors and makes the evaluation process more standardized.''} However, in this paper, we argue that subjectivity can play an important role in bias assessment in LLMs to augment the existing suite of automated evaluation methods. 
\section{Background}
\subsection{Bias Evaluation in LLMs}

Currently, LLMs are evaluated for social biases using automated or human evaluation methods. According to \citet{chang2023survey}, the former leverages metrics that can be \textit{``automatically calculated"}, while the latter requires human participation. We will briefly discuss both approaches in this section.

In automated evaluation of biases in LLMs, there are different processes and metrics available \cite{gallegos2023bias, chang2023survey, dhamala2021bold, parrish2021bbq, liang2023HELM, esiobu2023robbie}. For example, \textbf{Word, sentence or context embedding} \cite{caliskan2017semantics} are based on vector representations to identify biases in models. However, past work shows they can be unreliable in detecting biases in text \cite{cabello2023independence, goldfarb-tarrant-etal-2021-intrinsic}. Similarly, \citet{kaneko2022debiasing} consider \textbf{Token probability} (i.e., techniques that measure and compare the likelihood of a model's prediction under different conditions) can also be insufficient for bias evaluation and mitigation, especially in downstream tasks. Another popular bias detection approach involves prompting a model to then examine the text outputs. These \textbf{Generated text} techniques include, for instance, \textbf{Sentiment Analysis} of LLM text outputs, especially popular in the NLP community \cite{kiritchenko2018examining}; however, \citet{sheng2019babysitter} demonstrates that this approach might not be sufficient to detect more subtle biases in language. Next, it is essential to mention the multiple \textbf{prompting datasets and benchmarks} for bias evaluation. For example, adaptations of the  Winograd Schema \cite{levesque2012winograd}, such as WinoBias \cite{zhao2018winobias} and Winogender \cite{rudinger2018gender}; the latter offers datasets with sentences where gender is ambiguous to prompt an LLM and see which gendered pronoun is chosen; for example: \textit{The paramedic performed CPR on the passenger
even though she/he/they knew it was too late} \cite{rudinger2018gender}. This methodology assumes that the least-biased answer an LLM could give is the one closest to a predefined ground truth, for instance, the Census data showing if paramedics are mostly women or men. Other dataset examples include BOLD \cite{dhamala2021bold}, REalToxicityPrompts \cite{gehman2020realtoxicityprompts},
BBQ \cite{parrish2021bbq}, ROBBIE \cite{esiobu2023robbie} and HELM \cite{liang2023HELM} to cite a few. However, \citet{selvam2022tail, gallegos2023bias} alert to data reliability issues, indicating risks of these datasets and benchmarks being too static or even biased themselves (i.e., based on a chosen definition of what is bias in detriment to other opinions), and thus not equipped to deal with the ever-changing complexities and nuances of social biases. Besides, there are concerns around data contamination and benchmark leakage as part of an LLM's training dataset since most are widely available online \cite{zhou2023don, kotek2023gender}. This could result in \textit{``over-optimistic accuracy claims"} when applying these benchmarks \cite{liang2023HELM}.

The next cohort of bias evaluation methods for LLMs involves human participants. In order to detect such subtleties in social biases as described above, \textbf{human evaluation} can be leveraged. According to \citet{liang2021understanding}, human evaluators can detect more subtle cues and offer insights into clarity, coherence, and social fairness, which automated benchmarks might overlook. Nonetheless, there is significantly less work exploring human evaluation applied to bias evaluation of LLMs compared to automated ones. 
Existing work in LLM evaluation usually employs human evaluators mainly to verify results from the automated benchmarks \cite{huang2023bias, dhamala2021bold}. Interestingly, in some instances, human evaluations do not match with automatic evaluations \cite{esiobu2023robbie}, indicating a complex interplay between human appraisal and automated measurements. \citet{dhamala2021bold} uses human evaluation to compare with its automatic metrics and found that there was a lower correlation between human and machine evaluations when it comes to measuring toxicity and sentiment, potentially because these aspects \textit{``more strongly depend on the textual context which humans can more easily identify than classifiers."}
Further understanding this misalignment is critical to a more comprehensive bias evaluation of LLMs and potentially opening avenues for complementary performance in Human-AI Collaboration \cite{donahue2022human, bansal2021most}, where the combined efforts result in better performance than each party would have reached individually.

\subsection{Approach and Case Study: Gender Biases in LLMs}
When considering bias evaluation in LLMs, gender bias (i.e. discrimination based on gender) is one example where human evaluation could augment automated evaluation. We will explore this potential in this section.

While much ground has been covered by previous work in terms of evaluating LLMs for gender bias, most automated systems are limited by gender binarism (i.e., men and women), for example, when assigning gendered pronouns to ambiguous sentences \cite{rudinger2018gender, sheng2019babysitter}. However, it is essential to note that any gender can be a victim of gender biases (also known as sexism), not only men and women but also transgender, non-binary and all gender identities. Further, there is more nuance in gender bias, as scholars have been calling attention to the prevalence of a subcategory of sexism, named \textbf{subtle sexism} \cite{Benokraitis1997}; it is used to describe instances of gender discrimination that are less overt (for example, not necessarily using slurs or derogatory language), perceived as accepted in society, or benevolent \cite{barreto2005burden} (a discriminatory situation that is seen as positive towards the victim).

In online contexts, such as social media, subtle sexism is not only more common than overt cases but is also more challenging to detect \cite{hall2016they, khurana2022hate}. Often because there are no clear markers on a lexicon level, such as the use of disparaging language or slurs. Some initiatives to implement automated hate speech detection, for example, disproportionately target drag queens and the LBGT community, flagging their content as toxic more often \cite{oliva2021fighting} because of their choice in words. Further, some marginalized communities use derogatory language as a form of resistance against oppression \cite{diaz-etal-2022-accounting}. Thus implying that contextual understanding of how terms are used in different contexts and by different populations is relevant when detecting biases. In fact, \citet{mitamura2017value} defines sexism assessment as subjective and dependent on an individual's values, experiences and beliefs. Consequently, what some might consider to be sexist, others might not. This creates an additional layer of complexity for automated detection of sexism in text. 

In AI-generated text, gender bias can result in toxic or stereotyped text outputs. Hence, multiple studies are dedicated to identifying this type of social bias in LLMs \cite{dhamala2021bold, caliskan2017semantics, sheng2019babysitter}. Most of the current methods propose automated evaluations based on a dataset to prompt an LLM and then measure if there are gender biases in the text outputs. More often than not, the existing approaches are gender-binary (e.g., men vs women), looking for imbalances in gender parity but ignoring diversity in the gender spectrum. Further, these techniques rely on ground truth metrics such as the Census breakdowns or annotated datasets, where an individual or group of individuals determine what is and what is not sexist \cite{garg2023handling}. However, as demonstrated by \citet{mitamura2017value}, a purely objective agreement is not always possible regarding  sexism, as some situations are open to interpretation, especially in nuanced, subtle sexism scenarios. In such cases, simply choosing just one interpretation as ground truth might result in a false sense that an LLM is unbiased when, in fact, it only responds to that single view and alienates other divergent opinions \cite{gallegos2023bias}.

Consequently, this paper argues for embracing subjectivity as a feature and not a flaw when analyzing gender biases in LLM outputs. Further, disagreements, either between human evaluators or between human and automated evaluations, are to be expected in subtle and open-to-interpretation cases of sexism (and potentially other social biases as well). Thus, we should incorporate and learn from these disagreements as they are inherent to subjective decision-making tasks \cite{davani2022dealing}. We discuss potential avenues to be explored in the next section.

\section{Future Paths}

The Natural Language Processing (NLP) community is the first place to draw inspiration from since NLP researchers have extensively worked on automated hate speech detection \cite{jahan2023systematic}. Interestingly, the issue of subjectivity when evaluating subtle sexism (and social biases) in LLMs described in the previous section is comparable to the problem of low human annotator agreement identified in NLP research. The latter is found especially when labelling training datasets involving ambiguous tasks \cite{akhtar2020modeling} such as identifying toxicity in language \cite{waseem-hovy-2016-hateful, waseem2016you}. \citet{gordon2022jury} presents an interesting solution to these \textit{irreconcilable disagreements about ground truth labels}; instead of going for the majority vote, in detriment to minority voices, they propose balancing for which opinions to take into consideration, depending on context. This initiative could serve as inspiration to add subjective perspectives of human evaluators in addition to automated evaluation methods when identifying biases in LLMs. Furthermore, discussions about multi-perspectives and data perspectivism \cite{cabitza2023toward} are also relevant and could inspire different approaches and metrics that value human subjectivity.

Another alternative for bias evaluation of LLMs stems from recent calls for bias management instead of aiming to completely debiasing an algorithm. \citet{ferrara2023should} highlights that as humans and human language exhibit social biases, it is expected that LLMs will reflect them as well; therefore wholly removing all biases might be an unfeasible task. In a bias management approach, as discussed in a recent article by \citet{demartini2023data}, biases are not entirely removed but instead surfaced to end-users, leveraging Explainable AI (XAI) techniques \cite{liao2021human}. One potential avenue to be explored is leveraging human evaluation when end-users interact with a Large Language Model (LLM), similar to the one proposed by \citet{shen2021everyday}. For instance, since sexism is inherently subjective and value-based (some end-users might consider an AI output as sexist, while others might not), embracing the subjective nature of gender bias should also be part of designing an LLM evaluation system. Hence, instead of aiming to calibrate the algorithm to an imperfect proxy ground truth (e.g., an annotated dataset with one single definition of what is or is not sexist), the system could elicit feedback from end-users in-situ to employ cultural adaptations as needed \cite{johansen2020studying, reinecke2011improving, lum2022biasing}. In practice, this could be formatted as user surveys while the end-user is interacting with the LLM to evaluate its outputs for biases. A design inspiration could be the existing spelling and grammar checking tools offered by Microsoft and Grammarly that also nudge users to use more inclusive language but then adding an interactive component for user feedback in terms of biases in the recommendation. This more human-centric evaluation approach also aligns with \citet{zhu2018value}'s concept of value-sensitive algorithm design, advocating for pluralism in stakeholder needs that systems must meet. Additionally, there is further a potential for human-AI collaboration if the system incorporates user feedback iteratively.

\section{Conclusion}

In this paper, we argue that automated techniques to evaluate if an LLM is biased represent a significant advance towards more responsible AI applications. Nonetheless, as discussed in this work, there is an opportunity to explore human evaluation and embrace subjectivity for a more holistic comprehension of the representation of social biases in these models and how they impact human-AI collaboration.

 \pdfoutput=1
 

\bibliographystyle{ACM-Reference-Format}



\end{document}